\documentclass[final,5p,times,twocolumn]{elsarticle}
\usepackage{lineno,hyperref,amsmath,amsthm,amssymb,amsfonts,ragged2e,color}
\journal{``High Energy Density Physics"}
\begin{document}
\begin{frontmatter}
\title{Generation of rogue waves in space dusty plasmas}
\author{M.H. Rahman$^*$, A. Mannan, N.A. Chowdhury, and A.A. Mamun}
\address{Department of Physics, Jahangirnagar University, Savar, Dhaka-1342, Bangladesh\\
Email: $^*$rahman1992phy@gmail.com}
\begin{abstract}
  The basic features of dust-acoustic (DA) waves (DAWs) in four component dusty plasma
system (containing inertial cold and hot dust grains, inertialess non-extensive ions
and electrons) have been theoretically investigated by deriving the nonlinear
Schr\"{o}dinger equation. The analytic analysis under consideration demonstrates two
types of modes, namely, fast and slow DA modes. The unstable domain for the fast
DA mode, which can be recognized by the critical wave number ($k_c$), gives rise to the
DA rogue waves (DARWs). It is observed that the  amplitude and width of the DARWs
are significantly modified by various plasma parameters. The present results
should be useful in understanding the conditions for modulational instability of DAWs
and generation of DARWs in space dusty plasma systems like Saturn F-rings.
\end{abstract}
\end{frontmatter}
\section{Introduction}
\label{2sec:Introduction}
The observational data support the existence of massive charged dust grains not only in
astrophysical environments, viz. F-rings of Saturn \cite{Selim2015},  Earth's mesosphere,
and Jupiter's magnetosphere \cite{Horanyi1993}, but also in many laboratory experiments, viz. ac discharge,
Q-machine, and rf discharges \cite{Shukla2002}, etc. The presence of highly charged massive
dust grains in plasmas can significantly modify the dynamics of the plasma medium.
Dust-acoustic (DA) waves (DAWs), in which mass density of dust grains provides moment of inertia
and thermal pressure of the ions and electrons provides restoring force to propagate DAWs,
have employed by the physicists to understand various nonlinear electrostatic structures, viz.
envelope \cite{Chowdhury2018} and rogue profile \cite{Selim2015}, in dusty plasmas (DP).

Sometimes, highly energetic inertialess particles in space and laboratory plasmas move very fast, due to external
force field and wave-particle interaction, compared to their thermal velocity. Such kind of highly energetic
inertialess particles are governed by the non-extensive $q-$distribution
function \cite{Tsallis1988,Chowdhury2017,Tasnim2015,Ferdousi2015,Saha2014,Amour2010,Emamuddin2013,Ghosh2012}. A number of authors have studied
various nonlinear waves in plasma medium by considering $q-$distributed inertialess plasma species.
Tasnim \textit{et al.} \cite{Tasnim2015} investigated  propagation of DA shock waves (DASHWs) in
$q-$distributed ionic plasma medium and the magnitude of the amplitude of DASHWs decreases with non-extensive parameter $q$.
Ferdousi \textit{et al.} \cite{Ferdousi2015} studied  DASHWs in presence of $q-$distributed ions and
found that the polarity and amplitude of the DASHWs depend on non-extensivity of ions.
Saha and Chatterjee \cite{Saha2014} reported that generation and propagation of DA solitary waves (DASWs)
in a two component DP with  $q-$distributed ions. Amour and Tribeche \cite{Amour2010} examined
the DASWs in a DP with $q-$distributed electrons and found that the non-extensivity of the electrons
makes the DASWs structure more spiky. Emamuddin \textit{et al.} \cite{Emamuddin2013} investigated DAWs
in a DP with ions and $q-$distributed electrons and observed that the amplitude of both positive and negative
Gardner solitons increases with non-extensivity. Ghosh \textit{et al.} \cite{Ghosh2012} studied the effect
of the non-extensivity of ions during the head-on collision of DASWs and  the phase shift in a DP
composed of dust and $q-$distributed ions.

The amplitude modulation of the nonlinear propagation in a dispersive media, due to
carrier wave self interaction or nonlinearity of the medium, is an intrigue mechanism.
The modulational instability (MI) of nonlinear propagation, which leads to generate
freak waves \cite{Kharif2009}, giant waves, rogue waves (RWs), and envelope solitons,
is governed by the nonlinear Schr\"{o}dinger equation (NLSE). The novelty of RWs have
attracted the attention of numerous researchers in various fields, viz. optics \cite{Solli2007},
super-fluid helium \cite{Ganshin2008}, hydrodynamics, stock market \cite{Yan2010},
and plasma physics \cite{Chowdhury2017}. For first time in 1977, Watanabe \cite{Watanabe1977}
experimentally observed the MI self-modulation of a nonlinear ion wave packet.
Subsequently, a number of theoretical investigations have been done to understand
the effect(s) of various plasma parameters, viz. plasma species temperature,
number density \cite{Selim2015}, charged state \cite{Selim2015}, and other factors,
on the MI characteristics of the nonlinear propagation in plasma medium. Bouzit and
Tribeche \cite{Bouzit2015} reported that the DA RWs (DARWs) structures are very
sensitive to any change in the restoring force acting on the dust particles.
El-Taibany and Kourakis \cite{El-Taibany2006} studied MI of DAWs in an unmagnetized warm DP medium
and  observed the effects of dust charge variation, dust temperature, and constituent
plasma particle concentration on the MI of DAWs. Moslem \textit{et al.} \cite{Moslem2011}
examined that the amplitude of the DARWs increases with the increase of $q$  in a non-extensive plasmas.
Selim \textit{et al.} \cite{Selim2015} investigated the propagation of nonlinear DARWs in presence
of cold and hot dust grains as well as iso-thermal electrons and non-thermal ions and found that non-thermal
ions decrease the nonlinearity of the plasma medium and amplitude of the DARWs. Bains \textit{et al.} \cite{Bains2013}
analyzed the MI of the DAWs in the presence of $q-$distributed electrons and ions and observed that
the instability occurs at higher value of $k$ with an addition of negative dust.
To the best of our knowledge, the effects of cold and hot dust as well as $q-$distributed electrons and ions
on the MI of DAWs and DARWs have not been investigated. Therefore, in our present investigation,
our aim is to examine the  MI of the DAWs, generation of the DARWs, and the effects
of the $q-$distributed electrons and ions on the DARWs in DP system composed of inertial
cold and hot ions and inertialess $q-$distributed electrons and ions.

The manuscript is organized as the following fashion: The basic model equations
are presented in Sec. \ref{2sec:Model Equations}. The MI is given in Sec. \ref{2sec:Stability of DAWs}.
Finally, a brief discussion is provided in Sec. \ref{2sec:Discussion}.
\section{Model Equations}
\label{2sec:Model Equations}
We consider an unmagnetized four component DP system which consists of inertialess $q-$distributed
electrons (charge $-e$; mass $m_e$; number density $n_e$) and ions (charge $+e$; mass $m_i$; number density $n_i$), inertial negatively
charged cold dust grains (charge $q_c=-eZ_c$; mass $m_c$; number density $n_c$) as well as negatively charged hot dust grains (charge
$q_h=-eZ_h$; mass $m_h$; number density $n_h$; adiabatic pressure $P_h$); where $Z_c$ ($Z_h$) is the charge state of the negatively charged cold (hot) dust grains.
The quasi-neutrality condition at equilibrium is $n_{i0}=n_{e0}+Z_c n_{c0}+Z_h n_{h0}$; where $n_{c0}$ ($n_{h0}$) is the number densities
of the negatively charged cold (hot) dust grains at equilibrium and $n_{h0}>n_{c0}$. The normalized governing equations of the system can be written as:
\begin{eqnarray}
&&\hspace*{-1.3cm}\frac{\partial n_c}{\partial t}+\frac{\partial}{\partial x}(n_c u_c)=0,
\label{2:eq1}\\
&&\hspace*{-1.3cm}\frac{\partial u_c}{\partial t}+u_c\frac{\partial u_c}{\partial x}=\frac{\partial \phi}{\partial x},
\label{2:eq2}\\
&&\hspace*{-1.3cm}\frac{\partial n_h}{\partial t}+\frac{\partial}{\partial x}(n_h u_h)=0,
\label{2:eq3}\\
&&\hspace*{-1.3cm}\frac{\partial u_h}{\partial t}+u_h\frac{\partial u_h}{\partial x}+ \delta n_h\frac{\partial n_h}{\partial x}=\sigma \frac{\partial \phi}{\partial x},
\label{2:eq4}\\
&&\hspace*{-1.3cm}\frac{\partial^2 \phi}{\partial x^2}=(\mu-1-\lambda)n_e-\mu n_i+n_c+\lambda n_h.
 \label{2:eq5}
\end{eqnarray}
The normalizing  parameters are defined as:
$n_c=N_c/n_{c0}$, $n_h=N_h/n_{h0}$, $u_c=U_c/C_{dc}$, $u_h=U_h/C_{dc}$,
$x=X/\lambda_{Ddc}$, $t=T\omega_{pdc}$, $\phi=e\tilde{\phi}/{k_B T_i}$, $C_{dc}=\sqrt{(Z_ck_BT_i/m_c)}$,
$\lambda_{Ddc}=\sqrt{(k_BT_i/4 \pi e^2 Z_c n_{c0})}$, $\omega_{pdc}=\sqrt{(4\pi e^2 Z_c^2 n_{c0}/m_c)}$,
$P_h=P_{h0}(N_h/n_{h0})^\gamma$, $P_{h0}=n_{h0}k_BT_h$, $\gamma=(N+2)/N$, $\sigma=Z_h m_c/Z_c m_h$,
$\lambda=Z_hn_{h0}/Z_cn_{c0}$, $\mu=n_{i0}/Z_c n_{c0}$, $\delta=3T_h m_c/Z_c T_i m_h$.
Where $N_c$, $N_h$, $U_c$, $U_h$, $X$, $T$, $\tilde{\phi}$, $k_B$, $C_{dc}$,  $\lambda_{Ddc}$,
$\omega_{pdc}$, $T_i$, $T_e$, $T_h$, and $P_{h0}$ is the dimensional number densities
of cold dust, hot dust, cold dust fluid speed, hot dust fluid speed, space co-ordinate, time co-ordinate,
electro-static wave potential, Boltzmann constant, sound speed of the negatively charged cold dust,
Debye length of the negatively charged cold dust grains, angular frequency of
the negatively charged cold dust, ions  temperature, electrons temperature, negatively charged
hot dust grains temperature, and the equilibrium adiabatic pressure of the negatively
charged cold dust grains, respectively. It may be noted here that we consider  $m_c=m_h$ and $T_i$, $T_e\gg T_h$.
$N$ is the degree of freedom and  for one-dimensional case $N=1$, hence $\gamma=3$.
The number density of the $q-$distributed \cite{Tsallis1988,Chowdhury2017} electrons
and ions can be given by the following normalized equation, respectively,
\begin{eqnarray}
&&\hspace*{-1.3cm}n_e= \left[1+\alpha(q-1)\phi\right]^{\frac{q+1}{2(q-1)}},
\label{2:eq6}\\
&&\hspace*{-1.3cm}n_i=\left[1-(q-1)\phi\right]^{\frac{q+1}{2(q-1)}},
\label{2:eq7}
\end{eqnarray}
where $\alpha=T_i/T_e$ ($T_e>T_i$). It may be noted here that (a) $q>1$ ($q<1$) stands for sub-extensive (super-extensive)
electrons and ions; (b) $q=1$ stands for Maxwellian electrons and ions. By substituting \eqref{2:eq6} and \eqref{2:eq7} into \eqref{2:eq5},
and expanding up to third order of $\phi$, we get
\begin{eqnarray}
&&\hspace*{-1.3cm}\frac{\partial^2 \phi}{\partial x^2}=-1-\lambda+n_{c}+\lambda n_{h}+\gamma_{1}\phi+\gamma_{2} \phi^{2}+\gamma_{3} \phi^{3}+\cdot\cdot\cdot,
\label{2:eq8}
\end{eqnarray}
where
\begin{eqnarray}
&&\hspace*{-1.3cm}\gamma_1=\frac{(q+1)(\alpha\mu-\alpha-\alpha\lambda+\mu)}{2},
\nonumber\\
&&\hspace*{-1.3cm}\gamma_2=\frac{(q+1)(q-3)(\alpha^{2}+\alpha^{2}\lambda+\mu-\alpha^{2}\mu)}{8},
\nonumber\\
&&\hspace*{-1.3cm}\gamma_3=\frac{(q+1)(q-3)(3q-5)(\alpha^{3}\mu-\alpha^{3}-\alpha^{3}\lambda+\mu)}{48}.
\nonumber\
\end{eqnarray}

To investigate the MI of the DAWs, we employ the reductive perturbation method to derive the appropriate NLSE. The independent
variables are stretched as $\xi=\epsilon(x-v_gt)$ and
$\tau=\epsilon^2t$, where $\epsilon$ is a small parameter and $v_g$ is the group velocity
of the wave. The dependent variables \cite{Chowdhury2018} can be expressed as:
\begin{eqnarray}
&&\hspace*{-1.3cm}\Lambda(x,t)=\Lambda_0+\sum_{m=1}^{\infty}\epsilon^{(m)}\sum_{l=-\infty}^{\infty}\Lambda_{l}^{(m)}(\xi,\tau)~\mbox{exp}(il\Upsilon),
\label{2:eq9}
\end{eqnarray}
where $\Lambda_{l}^{(m)}=[n_{cl}^{(m)}$, $u_{cl}^{(m)}$, $n_{hl}^{(m)}$, $u_{hl}^{(m)}$, $\phi_l^{(m)}]^T$,
$\Lambda_0=[1, 0, 1, 0, 0]^T$, $\Upsilon=(kx-\omega t)$, and $k$ ($\omega$) is the real variables presenting
the carrier wave number (frequency), respectively. We are going parallel as done in Chowdhury \textit{et al.} \cite{Chowdhury2018}
work to find successively dispersion relation, group velocity, and NLSE. The DAWs dispersion relation
\begin{eqnarray}
&&\hspace*{-1.3cm}\omega^2=\frac{k^2D\pm k^2 \sqrt{D^2-4ME}}{2M},
\label{2:eq10}
\end{eqnarray}
where $D=(1+\sigma\lambda+\delta\gamma_1+\delta k^2)$, $M=(\gamma_1+k^2)$, and $E=\delta k^2$.
In order to obtain real and positive values of $\omega$, the condition
$D^2>4ME$ must be satisfied and positive (negative) sign in \eqref{2:eq10}
is referred to fast DA mode $\omega_f$ (slow DA mode $\omega_s$). The group velocity $v_g$ of DAWs can be written as
\begin{eqnarray}
&&\hspace*{-1.3cm}v_g=\frac{2\omega S^2-2S^2\omega^3+\sigma \lambda \delta k^2 \omega^3+\sigma \lambda \omega^5-\sigma \lambda S\omega^3}{2(kS^2+\sigma\lambda k \omega^4)},
\label{2:eq11}
\end{eqnarray}
where $S=\delta k^2-\omega^2$. Finally, the following NLSE:
\begin{eqnarray}
&&\hspace*{-1.3cm}i\frac{\partial \Phi}{\partial \tau}+P\frac{\partial^2 \Phi}{\partial \xi^2}+Q|\Phi|^2\Phi=0,
\label{2:eq12}
\end{eqnarray}
where $\Phi=\phi_1^{(1)}$ for simplicity and $P$ ($Q$) is the dispersion (nonlinear) coefficient, and is written by
\begin{eqnarray}
&&\hspace*{-1.3cm}P=\frac{F_1}{2 \omega Sk^2 (S^2+\lambda \sigma \omega^4)},
\nonumber\
\end{eqnarray}
\begin{eqnarray}
&&\hspace*{-1.3cm}Q=\frac{F_2}{2 k^2 (S^2+\lambda\sigma\omega^4)},
\nonumber\
\end{eqnarray}
where
\begin{eqnarray}
&&\hspace*{-1.3cm}F_1= S^3 (\omega-v_g k)(\omega-v_g k-2 \omega v_gk+2v_g^2 k^2)
\nonumber\\
&&\hspace*{-0.5cm}+\lambda \sigma \omega^4 (\delta k-\omega v_g) (2 \omega v_gk^2+kS-k \omega^2-\delta k^3)
\nonumber\\
&&\hspace*{-0.5cm}+(\omega-k v_g)(2v_gk \omega^2-\delta \omega k^2-\omega^3+v_gkS)-S^3 \omega^4,
\nonumber\\
&&\hspace*{-1.3cm}F_2=3 \gamma_3 S^2 \omega^3-\omega S^2 k^2 (A_1+A_6)-2 S^2 k^3 (A_2+A_7)
\nonumber\\
&&\hspace*{-0.5cm}-\sigma \lambda k^2 \omega^3 (\omega^2+\delta k^2) (A_3+A_8)-2 \lambda \sigma k^3 \omega^4 (A_4+A_9)
\nonumber\\
&&\hspace*{-0.5cm}+2 \gamma_2 S^2 \omega^3 (A_5+A_{10}),
\nonumber\\
&&\hspace*{-1.3cm}A_1=\frac{3 k^4-2 A_5 k^2\omega^2}{2 \omega^4},
\nonumber\\
&&\hspace*{-1.3cm}A_2=\frac{A_1 \omega^4-k^4}{k \omega^3},
\nonumber\\
&&\hspace*{-1.3cm}A_3=\frac{2 A_5 \sigma S^2 k^2-\delta \sigma^2 k^6-3 \sigma^2 \omega^2 k^4}{2 S^3},
\nonumber\\
&&\hspace*{-1.3cm}A_4=\frac{A_3\omega S^2-\omega \sigma^2 k^4}{k S^2},
\nonumber\\
&&\hspace*{-1.3cm}A_5=\frac{3 S^3 k^4+2 \gamma_2 S^3 \omega^4-3 \lambda \sigma^2 k^4 \omega^6-\delta \lambda
\sigma^2 \omega^4 k^6}{2 \omega^2 S^2 (S k^2-4 S \omega^2 k^2-\sigma \lambda \omega^2 k^2-\gamma_1 S \omega^2)},
\nonumber\\
&&\hspace*{-1.3cm}A_6=\frac{2 v_g k^3+\omega k^2-A_{10} \omega^3}{v_g^2 \omega^3},
\nonumber\\
&&\hspace*{-1.3cm}A_7=\frac{A_6 v_g \omega^3-2 k^3}{\omega^3},
\nonumber\\
&&\hspace*{-1.3cm}A_8=\frac{2 \omega v_g \sigma^2 k^3+\sigma^2 \omega^2 k^2+
\delta \sigma^2 k^4-\sigma A_{10} S^2}{S^2 (v_g^2-\delta)},
\nonumber\\
&&\hspace*{-1.3cm}A_9=\frac{A_8 v_g S^2-2 \omega \sigma^2 k^3}{S^2},
\nonumber\\
&&\hspace*{-1.3cm}A_{10}=\frac{F_3}{S^2 \omega^3 \left \{(v_g^2-\delta)+\lambda \sigma v_g^2-\gamma_1 v_g^2 (v_g^2-\delta)\right\}},
\nonumber\\
&&\hspace*{-1.3cm}F_3=2 \gamma_2 S^2 v_g^2 \omega^3 (v_g^2-\delta)+S^2 (v_g^2-\delta) (2 v_g k^3+\omega k^2)
\nonumber\\
&&\hspace*{-0.5cm}+\lambda v_g^2 \omega^3 (2 \omega v_g \sigma^2 k^3+\delta \sigma^2 k^4+\sigma^2 \omega^2 k^2)
\nonumber.
\end{eqnarray}
\begin{figure}[t!]
\centering
\includegraphics[width=70mm]{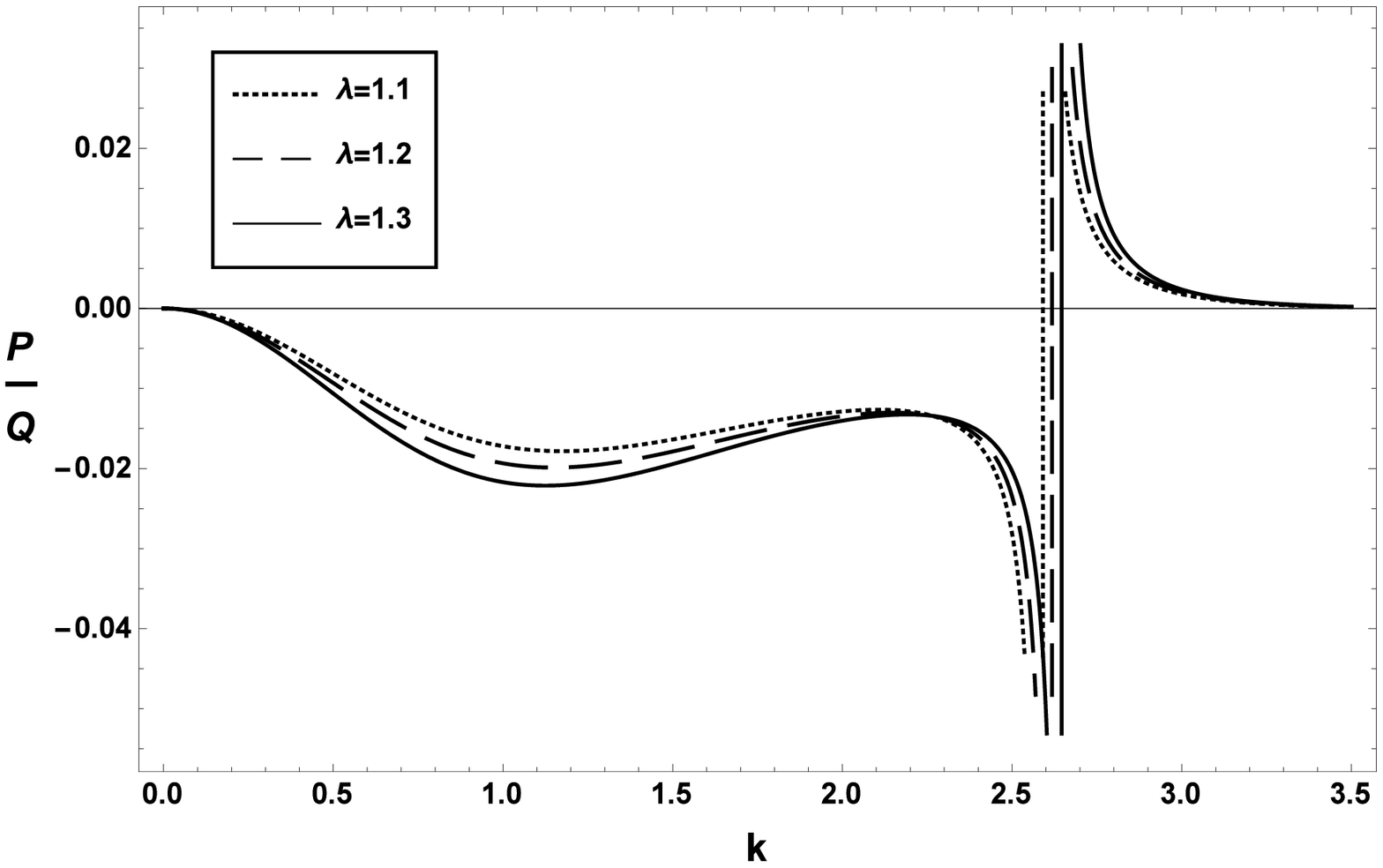}
\caption{The variation of $P/Q$ with $k$ for different values of $\lambda$;
along with fixed values of $\alpha=0.3$, $\delta=0.006$, $\mu=2.5$, $\sigma=0.5$, $q=1.8$ and $\omega_f$.}
\label{2Fig:F1}
\vspace{0.8cm}
\includegraphics[width=70mm]{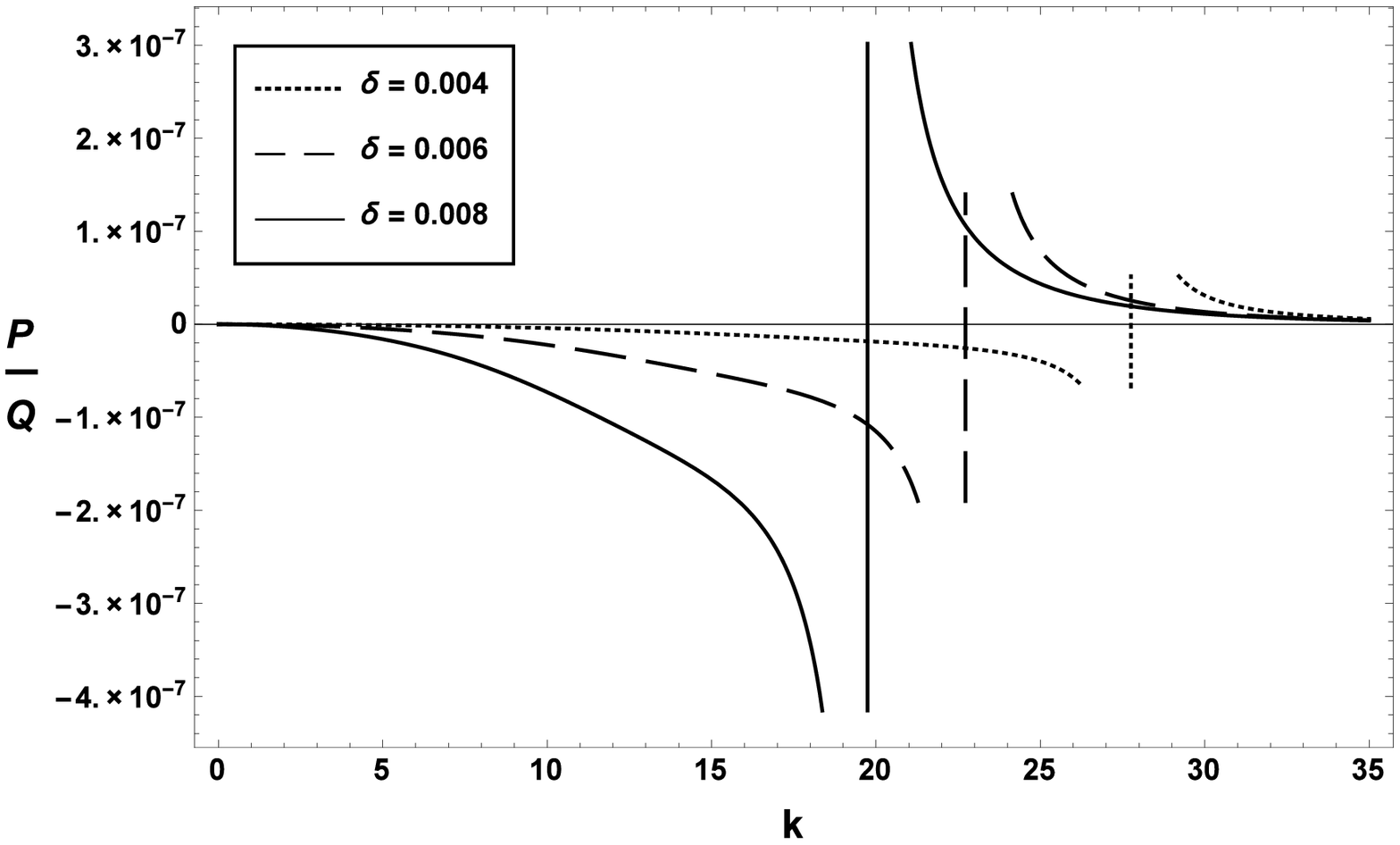}
\caption{The variation of $P/Q$ with  $k$ for different values of $\delta$;
along with fixed values of $\alpha=0.3$, $\lambda=1.2$, $\mu=2.5$, $\sigma=0.5$, $q=1.8$, and $\omega_s$.}
 \label{2Fig:F2}
\end{figure}
\section{Modulational instability}
\label{2sec:Stability of DAWs}
The stability of the DAWs depends on the sign of the nonlinear ($P$) and dispersive ($Q$)
coefficients \cite{Sultana2011,Schamel2002,Fedele2002,Kourakis2005}. Modulationaly stable
domain occurs for the DAWs when $P$ and $Q$ are opposite sign ($P/Q<0$). On the other hand,
modulationaly unstable domain occurs for the DAWs when $P$ and $Q$ are same sign ($P/Q>0$).
The point, in which $P/Q$ curve coincides with the $k-$axis in $P/Q$ vs $k$ graph, is known
as critical/threshold wave number ($k_c$) and this $k_c$ recognizes the stable/unstable
domain for the DAWs. The stability of the DAWs for the fast and slow DA modes can be
observed in Figs. \ref{2Fig:F1} and \ref{2Fig:F2}, respectively. It is obvious from
Fig. \ref{2Fig:F1} that the $k_c$ increases (decrease) with the increase of hot
dust concentration $n_{h0}$ (cold dust concentration $n_{c0}$) when  the charge state
of the cold ($Z_c$) and hot ($Z_h$) dust remain constant (via $\lambda$). Figure \ref{2Fig:F2}
shows that the stable domain of the DAWs increases with the increase in the value of
hot dust mass density ($m_h$), but decreases with increase of the cold dust mass
density ($m_c$) for constant value of hot dust temperature ($T_h$), ion temperature ($T_i$),
and cold dust charge state ($Z_c$), respectively, (via $\delta$).

The rogue waves solution of the NLSE \eqref{2:eq12} in the unstable domain, which
developed by Darboux Transformation Scheme, can be written as \cite{Akhmediev2009,Ankiewiez2009}:
\begin{eqnarray}
&&\hspace*{-1.3cm}\Phi(\xi,\tau)=\sqrt{\frac{2P}{Q}}\left[\frac{4(1+4iP\tau)}{1+16P^2\tau^2+4\xi^2}-1 \right]\mbox{exp}(2iP\tau).
\label{2:eq37}
\end{eqnarray}
\begin{figure}[t!]
\centering
\includegraphics[width=70mm]{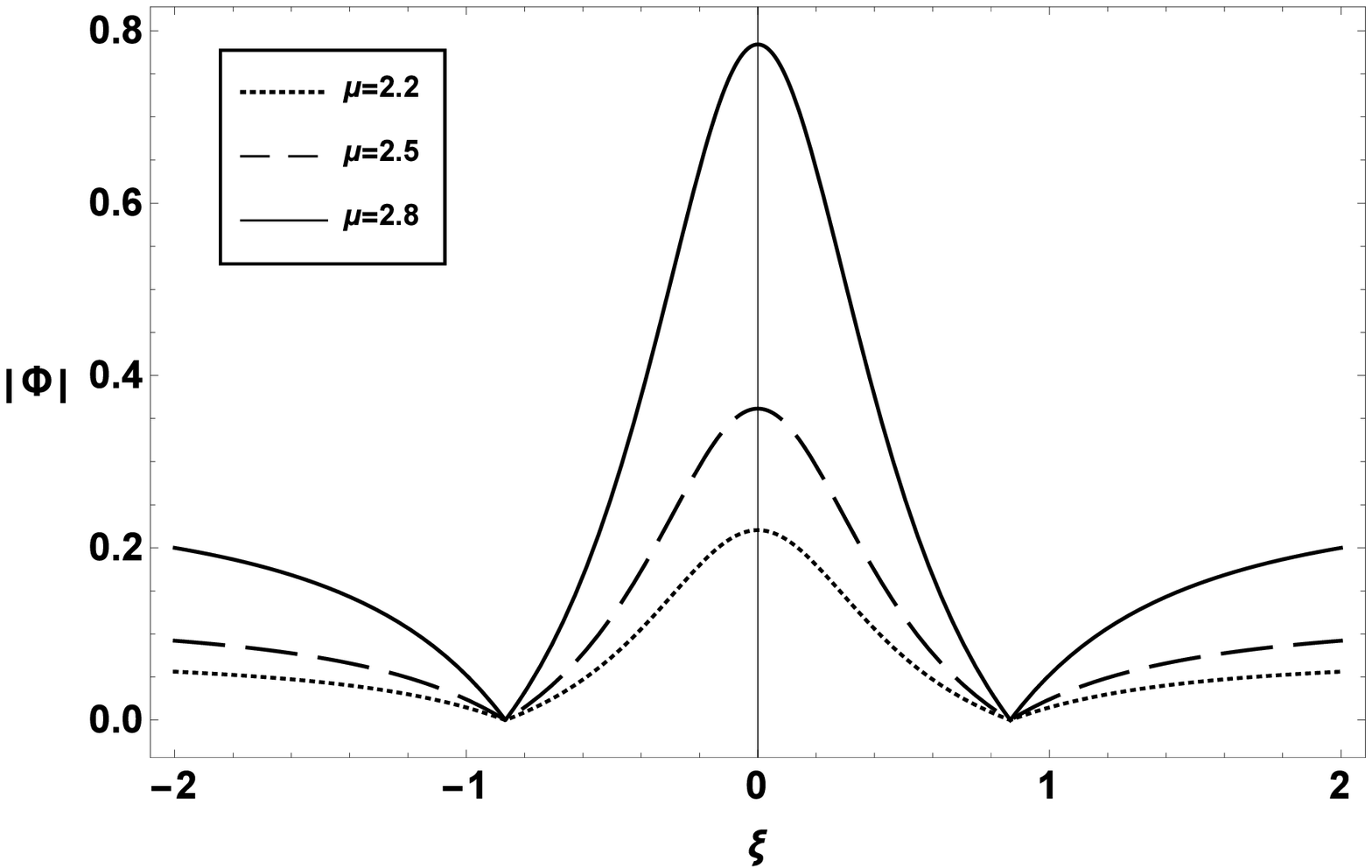}
\caption{The variation of $|\Phi|$ with $\xi$ for different values of $\mu$; along with fixed
values of $\alpha=0.3$, $\delta=0.006$, $\lambda=1.2$, $\sigma=0.5$, $k=2.8$, $\tau=0$, $q=1.8$, and $\omega_f$.}
 \label{2Fig:F3}
 \vspace{0.8cm}
\includegraphics[width=70mm]{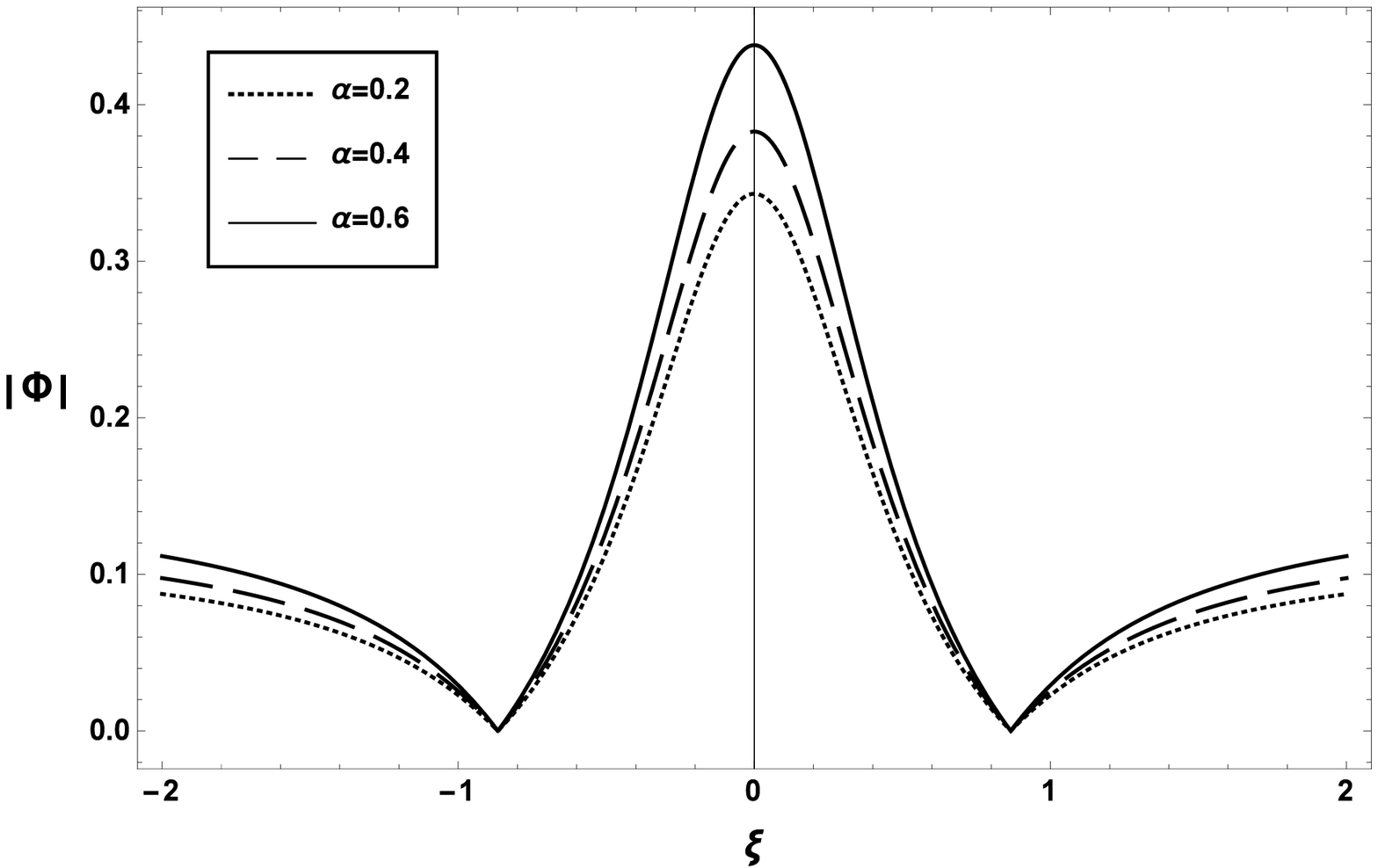}
\caption{The variation of $|\Phi|$ with $\xi$ for different values of $\alpha$; along with fixed
values of  $\delta=0.006$, $\lambda=1.2$, $\mu=2.5$, $\sigma=0.5$, $k=2.8$, $\tau=0$, $q=1.8$, and $\omega_f$.}
\label{2Fig:F4}
\end{figure}
\begin{figure}[t!]
\centering
\includegraphics[width=70mm]{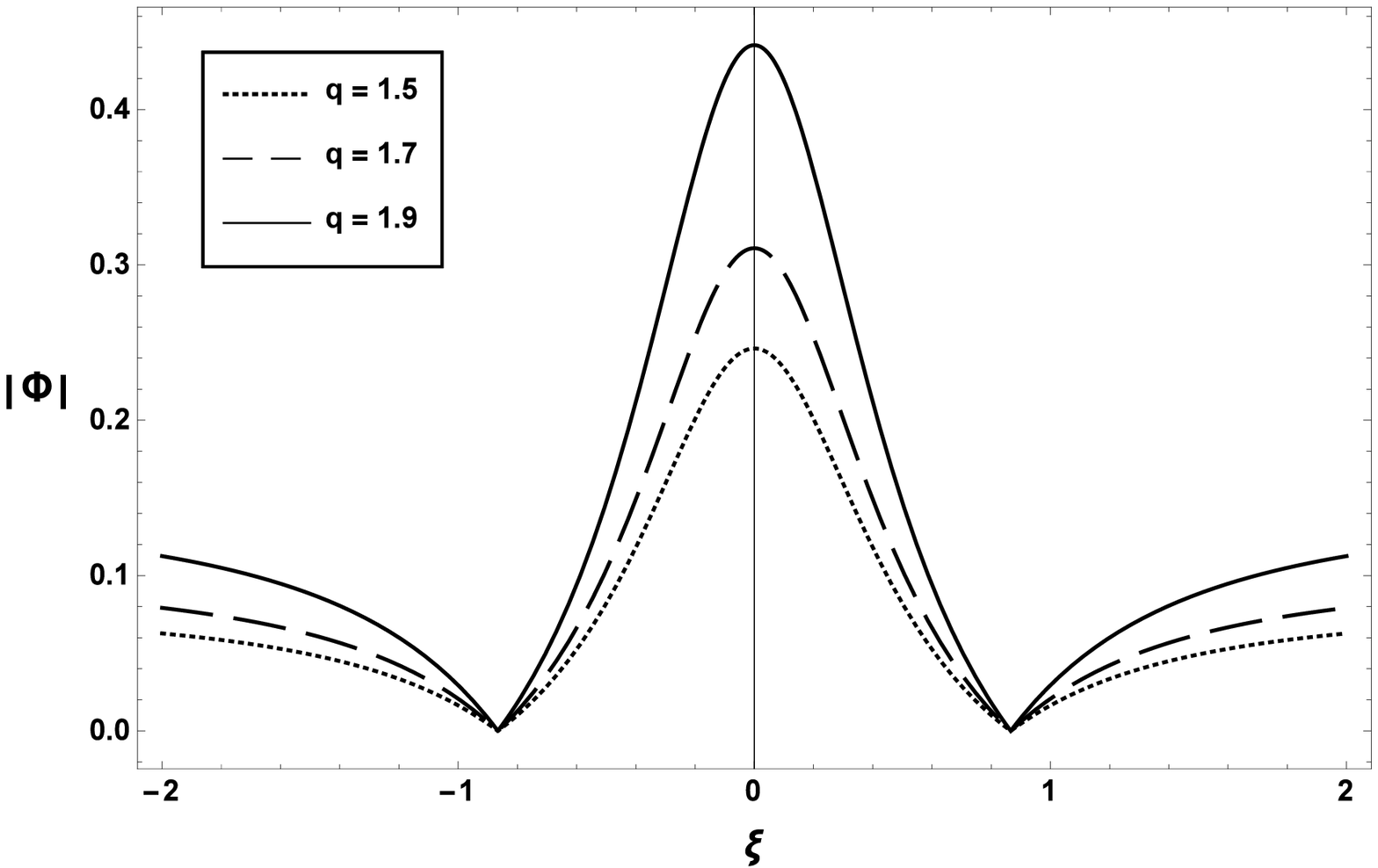}
\caption{The variation of $|\Phi|$ with $\xi$ for different values of positive $q$; along with fixed
values of  $\alpha=0.3$, $\delta=0.006$, $\lambda=1.2$, $\mu=2.5$, $\sigma=0.5$, $k=2.8$, $\tau=0$, and $\omega_f$.}
 \label{2Fig:F5}
 \vspace{0.8cm}
\includegraphics[width=70mm]{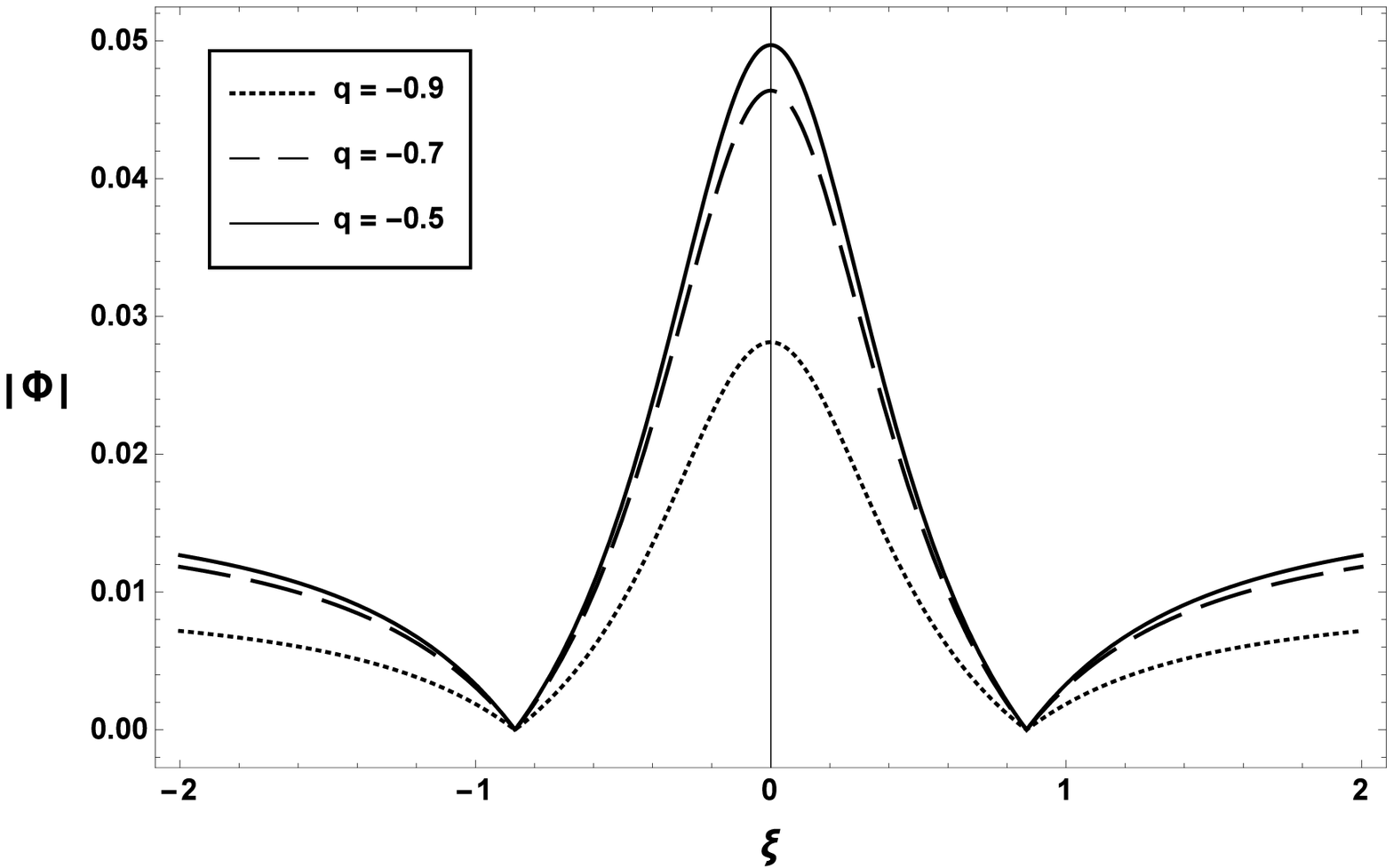}
\caption{The variation of $|\Phi|$ with $\xi$ for different values of negative $q$;
along with fixed values of $\alpha=0.3$, $\delta=0.006$, $\lambda=1.2$, $\mu=2.5$, $\sigma=0.5$,  $k=2.8$, $\tau=0$, and $\omega_f$.}
\label{2Fig:F6}
\end{figure}
The solution \eqref{2:eq37} implies that the concentration of high energy occurs (due
to the wave-particle interaction) within a small region. The effects of $q-$distributed
ion concentration on the shape of the DARWs can be observed from Fig. \ref{2Fig:F3}
and it is obvious that the amplitude of the DARWs increases with ion concentration $n_{i0}$ for
constant values of $Z_c$ and $n_{c0}$ (via $\mu$). Physically, the positively charged ions
enhance the nonlinearity of the plasma medium, and increase the amplitude of the electrostatic potentials.
So, the number density of $q-$distributed ion plays a vital role to control the shape of the DARWs.

The DARWs are so much sensitive to change in the values of ion temperature ($T_i$)
and electron temperature ($T_e$). It can be shown from Fig. \ref{2Fig:F4} that
the height of the DARWs increases (decreases) with ion (electron) temperature.
The physics of this result is that the nonlinearity of the plasma medium enhances with
ion temperature, that leads to generate a high energetic DARWs.

The effects of electrons and ions non-extensivity can be observed
from Figs. \ref{2Fig:F5} and \ref{2Fig:F6} and it is obvious  that (a) the
amplitude and width of DARWs increase with an increase $q$ (for both $q>0$ and $q<0$) and
the physics of this result is that the nonlinearity of the plasma medium increases with $q$;
(b) the amplitude of DARWs is independent to the sign of $q$, but dependent on the
magnitude of $q$ and this is a good agreement with Chowdhury \textit{et al.} \cite{Chowdhury2017} work;
(c) a comparison between DARWs potential for $q>0$ and $q<0$ can be observed from these two figures;
(d) the magnitude of the electrostatic potential for same interval of positive $q$ is not equal as negative $q$.
\section{Discussion}
\label{2sec:Discussion}
We have investigated the MI of a four component realistic DP medium by using standard NLSE.
The nonlinear and dispersive coefficients of the NLSE can be recognized the stability of the DAWs for fast and slow DA modes.
The $k_c$ value, which determines the stability conditions of DARWs, totally depends on dust masses, charge
state of dusts, and number density of the cold and hot dust. The core results from our present investigation
can be summarized as follows:
\begin{enumerate}
\item{Both $\omega_f$ and $\omega_s$ admit modulationally stable and unstable domain for DAWs.}
\item{The stable domain of the DAWs increases with the increase in the value of $m_h$, but decreases
with $m_c$ for constant value of $T_h$, $T_i$, and $Z_c$ (via $\delta$).}
\item{The amplitude of the electrostatic rogue profile decreases with ion concentration for constant values of $Z_c$ and $n_{c0}$ (via $\mu$).}
\item{The amplitude and width of DARWs increase with an increase $q$ (for both $q>0$ and $q<0$).}
\item{The amplitude of DARWs is independent to the sign of $q$, but dependent on the magnitude of $q$.}
\end{enumerate}
The present results may help in understanding the conditions of MI of DAWs and generation of DARWs
in four component space DP system, viz. F-rings of Saturn.
\section*{Acknowledgement}
M. H. Rahman is grateful to the Bangladesh Ministry of Science and Technology for
awarding the National Science and Technology (NST) Fellowship.


\section*{References}
\begin{thebibliography}{99}
\bibitem{Selim2015} M. M. Selim, H. G. Abdelwahed, and M. A. El-Attafi, Astrophys Space Sci. \textbf{25}, 359 (2015).

\bibitem{Horanyi1993} M. Hor\'{a}nyi, G. E. Morfill, and E. Gr\"{u}n, Nature (London) \textbf{363}, 144 (1993).

\bibitem{Shukla2002} P. K. Shukla and A. A. Mamun, \textit{Introduction to Dusty Plasma Physics} (Institute of Physics, Bristol, 2002).

\bibitem{Chowdhury2018} N. A. Chowdhury, A. Mannan, M. M. Hasan, and A. A. Mamun, Vacuum \textbf{147}, 31 (2018).

\bibitem{Tasnim2015} I. Tasnim, M. M. Masud, M. G. M. Anowar, and A. A. Mamun, IEEE Trans. Plasma Sci. \textbf{23}, 2187 (2015).

\bibitem{Tsallis1988} C. Tsallis, J. Stat. Phys. \textbf{52}, 479 (1988).

\bibitem{Chowdhury2017} N. A. Chowdhury, A. Mannan, M. M. Hasan, and A. A. Mamun, Chaos {\bf27}, 093105 (2017).

\bibitem{Ferdousi2015} M. Ferdousi, M. R. Miah, S. Sultana, and A. A. Mamun, Astrophys Space Sci. \textbf{360}, 43 (2015).

\bibitem{Saha2014} A. Saha and P. Chatterjee, Astrophys Space Sci. \textbf{351}, 533 (2014).

\bibitem{Amour2010} R. Amour and M. Tribeche, Phys. Plasmas \textbf{17}, 063702 (2010).

\bibitem{Emamuddin2013} M. Emamuddin, S. Yasmin, and A. A. Mamun, Phys. Plasmas \textbf{20}, 043705 (2013).

\bibitem{Ghosh2012} U. N. Ghosh, P. Chatterjee, and S. Kumar Kundu, Astrophys Space Sci. \textbf{339}, 255 (2012).

\bibitem{Kharif2009} C. Kharif, E. Pelinovsky, and A. Slunyaev, Rogue Waves in the Ocean (Springer-Verlag, Berlin, 2009).

\bibitem{Ganshin2008} A. N. Ganshin, V. B. Efimov, G. V. Kolmakov, L. P. Mezhov-Deglin, and P. V. E. McClintock, Phys. Rev. Lett. \textbf{101}, 065303 (2008).

\bibitem{Solli2007} D. R. Solli, C. Ropers, P. Koonath, and B. Jalali, Nature \textbf{450}, 1054  (2007).

\bibitem{Yan2010} Z. Yan, Commun. Theor. Phys. \textbf{54}, 947 (2010).

\bibitem{Watanabe1977} S. Watanabe, J. Plasma Phys. \textbf{17}, 487 (1977).

\bibitem{Bouzit2015} O. Bouzit and M. Tribeche,  Phys. Plasmas \textbf{22}, 103703 (2015).

\bibitem{El-Taibany2006} W. F. El-Taibany and I. Kourakis, Phys. Plasmas \textbf{13}, 062302 (2006).

\bibitem{Moslem2011} W. M. Moslem, R. Sabry, S. K. El-Labany, and P. K. Shukla, Phys. Rev. E \textbf{84}, 066402 (2011).

\bibitem{Bains2013} A. S. Bains, M. Tribeche, and C. S. Ng, Astrophys Space Sci. \textbf{343}, 621 (2013).

\bibitem{Sultana2011} S. Sultana and I. Kourakis, Plasma Phys. Control. Fusion {\bf53}, 045003 (2011).

\bibitem{Schamel2002} R. Fedele and H. Schamel, Eur. Phys. J. B  \textbf{27}, 313 (2002).

\bibitem{Fedele2002} R. Fedele, Phys. Scr. \textbf{65}, 502 (2002).

\bibitem{Kourakis2005} I. Kourakis and P. K. Sukla, Nonlinear Proc. Geophys. \textbf{12}, 407 (2005).

\bibitem{Akhmediev2009} N. Akhmediev, A. Ankiewicz, and J. M. Soto-Crespo, Phys. Rev. E \textbf{80}, 026601 (2009).

\bibitem{Ankiewiez2009} A. Anikiewicz, N. Devine, and N. Akhmediev, Phys. Lett. A \textbf{373}, 3997 (2009).

\end{thebibliography}
\end{document}